\documentclass[twocolumn,footinbib]{revtex4}
\usepackage{amsmath,amssymb}
\usepackage{bm}% bold math

\newcommand\la{\left\langle}
\newcommand\ra{\right\rangle}
\newcommand\eq[1]{(\ref{#1})}

\newcommand\be{\begin{eqnarray}}
\newcommand\ee{\end{eqnarray}}
\newcommand\lb{\label}

\newcommand\dnetp{\delta N_{p-\bar p}}
\newcommand\dnetn{\delta N_{n-\bar n}}
\newcommand\dnetpi{\delta N_{\pi^+-\pi^-}}

\begin{document}

\title{
Proton number fluctuation as a signal of the QCD critical end-point
}

\author{Y. Hatta}
\affiliation{Department of Physics, Kyoto University, 
         Kyoto 606-8502, Japan\\
and
The Institute of Physical and Chemical Research (RIKEN), Wako, Saitama 351-0198, Japan}

\author{M.A. Stephanov}
\affiliation{Department of Physics, University of Illinois, Chicago, Illinois 60607-7059, USA \\
and
RIKEN-BNL Research Center, Brookhaven National Laboratory, Upton, New York 11973, USA
}

\date{January 2003}
\begin{abstract}
We argue that the event-by-event fluctuation of the proton number is a
meaningful and promising observable for the purpose of detecting 
the QCD critical
end-point in heavy-ion collision experiments. The long range
fluctuation of the order parameter induces a characteristic
correlation between protons which can be measured.
The proton fluctuation also manifests itself as anomalous
enhancement of charge fluctuations near the end-point, which
might be already seen in existing data.
\end{abstract}


\maketitle

The event-by-event fluctuations in heavy-ion collisions carry
information about the degrees of freedom of the created system 
and their correlations \cite{ga}. In particular, thermodynamic 
properties of QCD can be inferred from event-by-event
fluctuation measurements \cite{C_V,srs,srslong,jk,asakawa,jeon}.

Of particular interest are fluctuations originating from the
QCD critical end-point
\cite{barducci,berges,srs,srslong,berdnikov,fodor,hatta}.  Since the
fluctuation of the order parameter induces characteristic correlations
among particles, in particular pions, it is expected that the
end-point affects the event-by-event fluctuations of certain
observables in a nontrivial way
\cite{srs,srslong,misha}.

Here we discuss a new observable which may serve as a signal of
the end-point; the event-by-event fluctuation of the {\it net proton
number}, i.e., the number of the protons minus the number of
antiprotons observed \footnote{Similar conclusions apply
to the fluctuations of the number of protons or antiprotons
separately. The {\em net} proton number appears more
naturally in the theoretical analysis, and we focus on it for clarity.}.

Our starting point is the fact that the baryon number susceptibility
$\chi_B$ \cite{larry,kunihiro,go,gottlieb,bernard,gavai} 
diverges at the critical
end-point \cite{berges,srs,gavin,hatta}. 
$\chi_B$ is related to the average magnitude of the
fluctuation $\delta B$ of the baryon number:
\be
\chi_B = {1\over VT}\la (\delta B)^2\ra\quad
\ee
where $V$ and $T$ are the volume and the temperature.
 The divergence of $\chi_B$ is a consequence of the fact 
that the critical point is the end-point of a line of first
order phase transitions, which are characterized, in particular, by 
a jump in the baryon number density $B/V$
\footnote{Note that the rise of $\chi_B$ near $T_c$
at $\mu_B=0$ \cite{gottlieb,bernard} is not due to the
critical fluctuations, but to the liberation of QCD degrees of
freedom \cite{kunihiro,go,gavai}. 
In fact, $\chi_B$ is finite at $\mu_B=0$ and $T_c$.}.
% At the end-point the two
% phases on the opposite sides of the transition are fusing into one,
% becoming indistinguishable, thus making the fluctuations between the
% two thermodynamically inexpensive. Critical opalescense is the closest
% analog of this phenomenon. 
%Qualitative and quantitative aspects of the
%behavior of $\chi_B$ near the end-point have been investigated in~\cite{hatta}.

If, in a heavy-ion collision experiment, we could measure {\em all}
the baryons, the enhancement of the event-by-event fluctuation of the
baryon number in a given subvolume would be a signature of the
end-point. However, about one half of the emitted baryons are
undetected neutrons which certainly contribute to the fluctuation of
the baryon number. To what extent does the proton number fluctuation
alone reflect the divergence of $\chi_B$?

This letter is devoted to clarifying where, in the observed
quantities, the divergence occurs and advocating the proton number
fluctuation as a sensible and promising observable for the search of the
critical point in the heavy-ion experiments.

In this work, we confine ourselves to equilibrium thermodynamic
fluctuations. Various important issues such as the nonequilibrium
evolution of the fluctuations will be (and some already have been)
studied separately.

For simplicity and clarity we shall work in QCD with exact 
isospin invariance. The relevant corrections
due to isospin breaking are small as we discuss below. 
Let us first show that in this
case the isospin number susceptibility, $\chi_I$, is {\em finite} at the
end-point. The proof is based on the fact that the singular
behavior of thermodynamic quantities near the critical point
is due to the divergence of a certain correlation length. It is
the correlation length in the $\sigma$-channel, the channel with
quantum numbers of the chiral condensate $\la\bar\psi\psi\ra$ 
\cite{berges,srs}. A density-density correlator, such as 
$\chi_I= (1/T)\int d^3x \la V_0(x)V_0(0) \ra$ can diverge only if the
density can mix with the $\sigma$ field. For the isospin density this
mixing is strictly forbidden by the SU(2)$_V$ (isospin) symmetry. The
isospin density, $V_0(x)$, transforms as a triplet, $\bm 3$. On the
other hand, $\sigma$ is a singlet. The mixing is forbidden and there
is no singular contribution in $\chi_I$. \footnote{For a similar
reason the axial isospin susceptibility remains finite at $T=T_c$,
$\mu_B=0$ \cite{muia}.}

Small explicit breaking of the SU(2)$_V$ symmetry by the 
quark mass difference $m_u-m_d$ or the isospin chemical potential
$\mu_I$ 
%\footnote{See also ref. \cite{muI} for recent studies
%of the effect of $\mu_I$
%near the QCD end-point.
%In the models considered in [30] two fields 
%become massless at the end-point at $\mu_I=0$: isosinglet  
%$\la\bar uu+\bar dd\ra$ and isotriplet $\la\bar uu-\bar dd\ra$. 
%As a result, $\chi_I$ diverges, as can be seen from our symmetry
%argument.} 
will induce singularity in $\chi_I$, 
since $m_u-m_d$ and $\mu_I$ are SU(2)$_V$ triplets 
\footnote{In the sense that they couple to SU(2)$_V$ triplet fields.
The argument here is similar to \cite{kogut}.}, and can produce
$V_0\sigma$ mixing. In the context of heavy ion collisions
corresponding singular contributions are negligible.

%produces a singular part in the
%susceptibility suppressed by $(m_u-m_d)^2$ which we can neglect. 
%If we neglect quark masses altogether,
%the breaking due to small $\mu_I$ can produce mixing only with
%a {\rm pair} of $\sigma$ fields. The $V_0\sigma$ mixing is forbidden
%by the 
%O(4)$\sim$SU(2)$_V\times$SU(2)$_A$ symmetry of massless QCD.
%Under this symmetry $V_0$ and $\mu_I$ are sextets ${\bm 6}$
%\footnote{
%For this argument $\mu_I$ should be promoted to a field as in
%\cite{kogut}.}, while 
%$\sigma$ is a quartet $\bm 4$. There are no singlets in $\bm
%{6\times6\times4}$, only in $\bm {6\times6\times4\times4}$. This means
%that, in addition to being proportional to $\mu_I^2$, the divergence
%will be controlled by the exponent $\alpha$ which is much
%smaller than the exponent $\gamma$ of the divergence of $\chi_B$.
%With finite quark masses there is a $V_0\sigma$ mixing proportional to 
%$(m_u+m_d)\mu_I$.

We summarize by writing the singular parts of the baryon and isospin
number susceptibilities:
\be
\chi_B\sim \xi^{\gamma/\nu}, \quad \chi_I=0 
\quad\mbox{(singular parts only)},
\lb{chichi}
\ee
where $\xi$ is the divergent correlation length of the sigma field: 
$\xi=1/m_\sigma$. The zero in eq. \eq{chichi}
neglects small isospin breaking terms as well as finite terms. 
The universal values of the exponents are given by
$\gamma\approx1.2$, $\nu\approx 0.63$, $\alpha\approx0.12\ll\gamma$.
Note that $\gamma/\nu=2-\eta$, where $\eta\approx0.04$.
% is the anomalous
%dimension of the $\sigma$-field.

Let us discuss the implications of \eq{chichi}. In particular,
let us consider charge susceptibility. Study of charge
fluctuations in heavy-ion collisions has attracted much attention
recently. It was proposed that these fluctuations might reflect
thermodynamic conditions earlier in the collision history, due to
charge conservation \cite{asakawa,jeon}. The charge
fluctuation magnitude per entropy is a measure of the charge per
particle or degree of freedom. 
In the quark-gluon plasma (QGP) the charge per degree of freedom is smaller.
If the charge fluctuation is equilibrating too slowly, the observed
value will be {\em smaller} than the equilibrium magnitude
which can be calculated using the resonance gas \cite{jk}.
However, the estimates of the
charge diffusion \cite{shuryak} suggest that for the
rapidity windows achievable in present experiments 
charge diffusion is very efficient in equilibrating charge
fluctuations, thus practically washing out the ``history'' effects.
Experimentally, the QGP suppression of the charge fluctuation is not seen
\cite{reid,spsdata}, which is consistent with the diffusion estimates
\cite{shuryak}. The effect of the critical fluctuations we are
discussing here is crucially different from the QGP suppression.
While the latter is the ``history'' effect, the critical fluctuations
are the {\em equilibrium} fluctuations pertaining to the freezeout
 point, and the diffusion is necessary to establish
them.

The measure of charge fluctuations, the 
charge number susceptibility, $\chi_Q$, can be expressed in terms
of $\chi_B$ and $\chi_I$ using the relation $Q=B/2+I_3$ and the
fact that isospin symmetry requires $\la \delta B\,\delta I_3 \ra=0$:
\be
\chi_Q = \frac1{VT}\la (\delta Q)^2 \ra= {1\over4} \chi_B + \chi_I.
\ee
Eq. \eq{chichi}  then implies that the charge susceptibility
diverges at the critical point, due to the divergence of $\chi_B$.

We now wish to relate the susceptibilities $\chi_B$, $\chi_I$ and
$\chi_Q$ to observable particle number fluctuations. For simplicity, 
we shall limit
our discussion by considering only protons, neutrons and pions.
Accounting for other particles will not alter our conclusions.
In the hadron language, the susceptibilities may be written as
\begin{eqnarray}
\chi_B&=&\frac{1}{VT}\la \left(
\delta N_{p-\bar p}+\delta N_{n-\bar n}
\right)^2 \ra \nonumber \\
\chi_I&=&\frac{1}{VT}\la\left(
\frac12\delta N_{p-\bar p}-\frac12\delta N_{n-\bar n}+\delta N_{\pi^+-\pi^-}
\right)^2 \ra
\end{eqnarray}
and
\begin{eqnarray}
%\nonumber \\
\chi_Q&=&\frac{1}{VT}\la \left(
\delta N_{p-\bar p}+\delta N_{\pi^+-\pi^-}
\right)^2 \ra 
\end{eqnarray}
where we introduced notation $N_{p-\bar p} \equiv N_{p} -
N_{\bar p}$ for the net proton number fluctuation, with $\delta$
denoting event-by-event deviation from the equilibrium value. Similar
notations are used for neutrons and pions.

Now we concentrate on singular parts of the susceptibilities and ask a
question: what does \eq{chichi} imply for the individual 
particle number fluctuations? It is easy to check that the following
set of relations between singular parts of the particle correlators
reproduces the correct singular behavior given in \eq{chichi}:

\begin{eqnarray}
&&\la\dnetp\,\dnetp\ra=\la\dnetn\dnetn\ra=\la\dnetp\,\dnetn\ra
\nonumber\\
&&\la\dnetpi\dnetpi\ra=0
\nonumber\\
&&\la\dnetp\,\dnetpi\ra=\la\dnetn\dnetpi\ra=0
\nonumber\\
&& \qquad\mbox{(singular parts only)}.
%
% \langle \delta N_p \delta N_p\rangle&=&\langle\delta N_p \delta N_n\rangle=\langle\delta N_n \delta N_n\rangle  \nonumber \\
% \langle\delta N_{\pi^+} \delta N_{\pi^+}\rangle&=&\langle\delta N_{\pi^+} \delta N_{\pi^-}\rangle=\langle\delta N_{\pi^-} \delta N_{\pi^-}\rangle\nonumber \\
% \langle\delta N_p \delta N_{\pi^+}\rangle&=&\langle\delta N_p \delta N_{\pi^-}\rangle \nonumber \\
% \langle\delta N_n \delta N_{\pi^+}\rangle&=&\langle\delta N_n \delta N_{\pi^-}\rangle \nonumber \\
% &&\mbox{(only the singular part)} 
\lb{k}
\end{eqnarray}
Some of these equations follow trivially from isospin invariance,
but some, for instance, the last equation on the first line and that on
the second line, require a stronger condition. Such relations occur
naturally if we attribute the divergences to the exchange of a sigma
meson, which is an isospin singlet. 
Using eqs. \eq{k} we obtain
\begin{eqnarray}
\chi_B&=&\frac{4}{VT}\la \dnetp\, \dnetp\ra,\nonumber \\
\chi_I&=&0,\nonumber \\
\chi_Q&=&\frac{1}{VT}\la \dnetp\, \dnetp\ra. \nonumber \\
&&\mbox{(singular parts only)}\lb{sin}
\end{eqnarray}
Remarkably, the singular part of the charge fluctuation comes from the
protons. In other words, had we considered only contributions
from charged pions in $\chi_Q$, the singular parts of
$\pi^+\pi^+$, $\pi^-\pi^-$, $\pi^+\pi^-$ correlators 
%$\la
%\delta N_{\pi^+}\delta N_{\pi^+}\ra$, $\la
%\delta N_{\pi^-}\delta N_{\pi^-}\ra$, $\la
%\delta N_{\pi^+}\delta N_{\pi^-}\ra$, 
(all are singular at the critical point \cite{srslong}) 
would have canceled each other.  We see also that the proton
number fluctuation completely reflects the singularity of the
baryon number susceptibility, which justifies its use 
as a sensible probe of the QCD critical end-point.

To provide a simple estimate of how large the net 
proton number fluctuation can become near the critical point,
we begin by calculating the correlator 
\begin{eqnarray}
\langle \delta n_{\bm p}\delta n_{\bm k}\rangle,\label{bin}
\end{eqnarray}
where $n_{\bm p}$ is the net proton number in the momentum bin labeled
by the value ${\bm p}$. In addition to the usual statistical
fluctuation, the correlator (\ref{bin}) receives a contribution from
the effective interaction with the sigma field $\sigma$,
%\begin{eqnarray}
${\cal L}_{\sigma pp}=g\sigma \bar{P}P$,
%\end{eqnarray}
where $g$ is the dimensionless sigma-nucleon coupling and $P$
is the Dirac field of a proton. 
All fluctuation observables of the protons can be constructed from 
(\ref{bin}) \cite{srs,misha}.
 
Near the critical point, the singular term in (\ref{bin}) is
represented by a diagram of forward proton-proton scattering.
%\cite{amp}.  
A straightforward
calculation following \cite{misha} gives,
\begin{eqnarray}
V\langle \delta n_{\bm p}\delta n_{\bm k}\rangle&=&
%2(2\pi)^3 \delta ({\bf p-k}) (n_p^+(1-n_p^+)+n_p^-(1-n_p^-)) \nonumber \\
%&+&
\frac{g^2}{m_{\sigma}^2T}\frac{4m^2}{E_{\bm p} E_{\bm k}}\bigl[n_{\bm p}^+(1-n_{\bm p}^+)-n_{\bm p}^-(1-n_{\bm p}^-)\bigr] \nonumber \\
&\times& \bigl[n_{\bm k}^+(1-n_{\bm k}^+)-n_{\bm k}^-(1-n_{\bm
k}^-)\bigr] \label{m}
\nonumber\\
& &\qquad\mbox{(singular part only)}
\lb{corr}
\end{eqnarray}
where $m=940$ MeV is the proton mass, $E_{\bm p}=\sqrt{{\bm p}^2+m^2}$ and
%$\displaystyle{n^{\pm}_{\bm p}=\left[\exp\left\{\frac{E_{\bm p}\mp
%\mu_B}{T}\right\}+1\right]^{-1}}$,
%\begin{eqnarray}
$n^{\pm}_{\bm p}=\left[\exp\left\{(E_{\bm p}\mp
\mu_B)/{T}\right\}+1\right]^{-1}$,
%\end{eqnarray}
while
$m_{\sigma}=1/\xi$ is the sigma meson
(screening) mass. 

Let us compare the singularity in \eq{corr}
to the singularity in \eq{chichi}. The exponent $\gamma/\nu=2-\eta$ 
in \eq{chichi} is very close to 2 (the anomalous dimension $\eta$
of the $\sigma$ field is small)
% (this can be traced to the smallness of exponent $\eta$),
and is equal to 2 in the mean field approximation ($\eta=0$), 
which is the same as the power of $1/m_\sigma$ in~\eq{corr}.

%% The behavior of the mass $m_\sigma$ on the QCD phase diagram near the
%% critical end point is as follows \cite{srs,allton}. $m_\sigma$
%%  is small (but
%% nonzero) on the crossover line due to the proximity of the second
%% order phase transition line in the chiral limit. $m_\sigma$
%% decreases as one approaches the critical point, where it vanishes %.
%% as $|\ell|^\nu\approx |\ell|^{0.63}$ 
%% if one follows the crossover line, where $|\ell|$
%% is the distance from the critical point on the $T\mu$ phase diagram.
%% It vanishes steeper if one approaches the critical point from
%% any other direction, as $|\ell|^{\nu/(\beta\delta)}\approx
%% |\ell|^{0.41}$.
%
In a realistic heavy-ion collision environment finiteness of 
the space-time volume severely prevents $m_\sigma$
from vanishing exactly \cite{srslong,berdnikov}. The smallest
achievable value is estimated to be around $(3\mbox{ fm})^{-1}$.

One possible concern is that the rescattering in the final hadronic
stage washes out critical point fluctuations. In this respect, one
should bear in mind that the rescattering in question includes the
exchange of the $\sigma$ quanta, which, near the critical point, is
the source of the fluctuations we consider.  The critical
fluctuations are washed out if the final (kinetic) freezeout occurs
sufficiently far from the critical point.  In order to
see the effect, one should dial control parameters (e.g., reduce the
size of the ions to raise the freezeout temperature) to bring the
freezeout closer to the critical point.

In (\ref{m}), $g$ is taken at zero momentum transfer, i.e., off the
sigma mass shell. In vacuum, $g\simeq m/f_{\pi}$ ($f_{\pi}=93$ MeV is
the pion decay constant) is quite large $\sim$10.  For the
quantitative estimate below, we assume that $g$ does not change
appreciably from its vacuum value near the chiral phase
transition. (See, however, \cite{hatsuda}.)

First, let us assume that Au-Au collisions at RHIC at $\sqrt{s}=130$
GeV froze out in the vicinity of the end-point and estimate the effect
of \eq{corr} in terms of the unknown mass $m_\sigma$, which is the
measure of the proximity of the end-point.  At the chemical freezeout,
$T=174$ MeV and $\mu_B=46$ MeV  \cite{braun}.  Integrating
over ${\bm p}$ and ${\bm k}$, we obtain the net proton number
fluctuation $\langle (\delta N_{p-\bar p})^2\rangle$ and divide it by the
{\it sum} of proton and antiproton numbers ${\langle N_{p+\bar{p}}\rangle}$:
\begin{eqnarray}
\frac{\langle (\delta N_{p-\bar p})^2\rangle}{\langle N_{p+\bar{p}}
\rangle}\bigg|_{\rm RHIC}
\approx 1.0+0.062\Bigl(\frac{g}{10}\Bigr)^2 \Bigl(\frac{200 \mbox{MeV}}{m_{\sigma}}\Bigr)^2. \label{f}
\end{eqnarray}
The unity on the r.h.s. is the trivial statistical contribution.
Taking $g=10$ and $m_{\sigma}=60$ MeV $\approx (3 \mbox{ fm})^{-1}$
\cite{srslong,berdnikov}, we find $
%\begin{eqnarray}
%\frac
{\langle (\delta N_{p-\bar p})^2\rangle}/{\langle N_{p+\bar{p}}\rangle }\approx 1.7
%\end{eqnarray}
$.

On the other hand, if the end-point were located at a value of $\mu_B$
of order of a few hundred MeV, as
inferred from simple model estimates \cite{berges}  and suggested by 
the recent lattice simulation \cite{fodor}, it is possible that the
SPS freezout is in the proximity of the critical point.
Using the freezeout parameters ($T, \mu_B$)=(168 MeV, 266 MeV) at 
SPS \cite{sps} 
%and ($T, \mu_B$)=(120 MeV, 540 MeV) at AGS \cite{ags}, 
we obtain
\begin{eqnarray}
\frac{\langle (\delta N_{p-\bar p})^2\rangle}{\langle N_{p+\bar{p}}
\rangle}\bigg|_{\rm SPS}
\approx 1.0+1.5\Bigl(\frac{g}{10}\Bigr)^2\Bigl(\frac{200 \mbox{MeV}}{m_{\sigma}}\Bigr)^2. \label{sp}
\end{eqnarray}
%for SPS, and
%\begin{eqnarray}
%\frac{\langle (\delta N_p)^2\rangle}{\langle N_{p+\bar{p}} \rangle}\approx 1.0+3.0\Bigl(\frac{g}{10}\Bigr)^2\Bigl(\frac{200 \mbox{MeV}}{m_{\sigma}}\Bigr)^2. \label{ag}
%\end{eqnarray}
%for AGS.
Note that the coefficient of the second term has a much bigger value
than in (\ref{f}). This is because the singular term given by (\ref{m})
grows as the square of the net proton number and also because 
there is a partial
cancellation between protons and antiprotons at RHIC. 
We stress that the main feature in (\ref{f}) and (\ref{sp})
is the singular dependence on $m_\sigma$, which makes the effect
large when the freezeout occurs near the critical point.
There are other effects, which
contribute to the r.h.s. of (\ref{f}) and (\ref{sp}), but which are
not singular near the critical point (e.g., initial volume
fluctuations caused by impact parameter fluctuations).

Experimentally, separating protons and measuring proton fluctuations
is a feasible task in the RHIC as well as the SPS detectors.
We hope that such data analysis will be available soon. 

In order to test our ideas on the existing data we can,
using (\ref{f}) and (\ref{sp}),
estimate the contribution of the proton fluctuation to the
total charge fluctuation characterized by $D\equiv4\langle(\delta
Q)^2\rangle/\langle
N_{\rm tot}\rangle$ \cite{jeon}.
As \eq{sin} shows, pions do not contribute to the {\em singular}
 part of the charge fluctuation,
but they dilute such a contribution of the protons.
% If we include all charged hadrons as is usually done in experimental 
% measurements,
% the dynamical fluctuation does not change due to (\ref{k}) while the
% total multiplicity of charged particles $N_{\rm tot}=N_++N_-$, where
% $N_{+(-)}$ is the number of positively (negatively) charged partices,
%is about 10 times larger, due to predominance of pions. 
Using $\langle N_{\pi^++\pi^-}\rangle
\approx 10\langle N_{p+\bar{p}}\rangle$ and 
%
%\be
$
{\langle (\delta
N_{\pi^+-\pi^-})^2\rangle/\langle
N_{\pi^++\pi^-}\rangle} \approx 1-0.3 = 0.7,
$
%\ee
%
where the negative contribution $-0.3$ is due to the resonance decays
\cite{jk}, we obtain
\begin{eqnarray}
\frac{D}{4} \equiv \frac{\langle (\delta Q)^2 \rangle}{\langle
N_{\rm tot}\rangle }&\approx& \frac{\langle (\delta N_{p-\bar
p})^2\rangle+\langle (\delta N_{\pi^+-\pi^-})^2 \rangle}{\langle
N_{p+\bar{p}}\rangle +\langle N_{\pi^++\pi^-}\rangle}  \nonumber \\
&\approx&0.8, \label{d}
\end{eqnarray}
where we have neglected the cross terms between $\delta N_p$ and
$\delta N_{\pi^{\pm}}$.  We see that the fluctuation anomaly in the
proton sector can result in a larger charge fluctuation than the resonance
gas value $\approx 0.7$ by about 10$\%$. 
At SPS this effect is even stronger. At SPS, using (\ref{sp}), $g=10$
and $m_\sigma=200$ MeV we get
$\langle (\delta N_{p-\bar p})^2\rangle/\langle N_{p+\bar{p}}\rangle \approx
2.5$, and with $\langle N_{\pi^++\pi^-}\rangle
\approx 5\langle N_{p+\bar{p}}\rangle$ we obtain:  $D/4\approx 1$. 

Before these estimates can be compared to experiment, one must take into
account the effect of limited acceptance of a given detector.  It is easy
to see that this effect reduces deviations from $D=4$. Its estimates
range from
few percent corrections \cite{jeon} to almost complete elimination of
 deviations from $D=4$ \cite{zaranek}, depending on the assumptions
on the rapidity correlator of fluctuations and the width of the
acceptance window. We do not discuss these issues here, and
refer the reader to the literature. 

Experimentally, the data from RHIC
suggest that the magnitude of the fluctuation is slightly {\em
larger} than a thermodynamical fluctuation in a resonance gas \cite{reid}. This
effect is even more pronounced at SPS \cite{spsdata}. 
There are, of course, a number of possible explanations, for example,
(i) acceptance, as we have just discussed \cite{zaranek};
(ii) remnant initial state correlations;
(iii) decay of multiply charge clusters;
(iv) other nonequilibrium fluctuations 
(e.g., by a mechanism similar to \cite{gavin});
%In other words,
%the negative contribution from decays of resonances such as
%$\rho^0$ and $\omega$ must be at least partially canceled by some
%other yet unknown effect. One possibility is a
%non-equilibrium fluctuation (e.g., 
%by a mechanism similar to \cite{gavin}). 
In this letter we wish to point out that, quite independently of these
other effects, an {\em equilibrium} critical fluctuation due to the
proximity of the end-point could explain the enhancement of charge
fluctuations observed at RHIC and SPS. The independent measurement of
proton fluctuations that we suggested would be necessary to confirm and
sufficient to rule out this effect.

Is it possible that the light sigma effect is seen in both RHIC and
SPS experiments? If this happens, the region of $\mu_B$ where $m_\sigma$ is
small ($<200$ MeV) is rather wide -- of order hundred MeV. Although unlikely,
 this might not be
completely unnatural if one takes into account the fact that
$m_\sigma$ is suppressed on the crossover line stretching from the
end-point to $\mu_B=0$ axis, even though it vanishes only at the
end-point (see, e.g., Fig. 5 in \cite{hatta}).

As it should be clear from our discussion, measuring the charge fluctuations
 is not the most efficient way to search for the end-point,
although the effect may be seen in such observables too. 
%Our estimates also show that if the proton number fluctuation
%is as large as can be caused by the proximity of the critical point,
%it can explain the puzzling 
%charge fluctuation measurements at RHIC and SPS. 
%A smaller value of
%$m_\sigma$ is favored by the RHIC data which suggest that the critical
%point might be located in the region probed by RHIC. 
A direct measurement of the proton number
fluctuation as a function of the $\sqrt s$ 
of the collision is both feasible and is less afflicted by other
effects. Correlation of such a measurement 
with other proposed signatures of the critical
point (such as $p_t$ fluctuations \cite{srs}) would affirm the
discovery of the QCD critical point.

In conclusion, protons carry both the baryon and the electric
charges. They are sensitive to the fluctuation of the order
parameter. Due to the peculiar nature of the
end-point -- isospin blindness of the sigma field --
the singularity of the baryon number susceptibility is
completely reflected in the proton number fluctuation. Thus 
the net proton number fluctuation is a very useful observable. 
By studying the $\mu_B$ dependence of this fluctuation
one may discover and
determine the location of the critical point on the phase diagram
of QCD.

We are grateful to L. McLerran and R. Pisarski 
for many discussions as well as to M. Asakawa,
M. Kaneta and T. Kunihiro for comments.  Y. H. thanks RIKEN BNL Center
for hospitality during the completion of this work.  M.S. thanks RIKEN
BNL Center and U.S. Department of Energy [DE-AC02-98CH10886] for
providing facilities essential for the completion of this work. The
work of M.S.  is supported in part by DOE Outstanding Junior
Investigator Grant and by the Alfred P. Sloan Foundation.

%Furthermore, the event-by-event fluctuation of the transverse momentum
%of charged particles $\Phi_{pt}$ \cite{ga} observed by STAR (called
%$\Delta \sigma_{pt}$ in \cite{reid}) is unusually large:
%$\Phi_{pt}\approx 35 \mbox{MeV}$.  The value of $\Phi_{pt}$ of this
%magnitude is consistent with the effect of the light sigma on pion
%fluctuations
% estimated in \cite{srslong}.  
%$\Delta \sigma_{pt}$ or $\Phi_{pt}$ is an isospin ``blind'', 
%or charge independent,
%in the terminology of \cite{reid}, quantity. The measurement of a charge
%dependent analog of it was also reported: $\Delta
%\sigma_{\Delta}\approx 0$. The absence of the charge dependent
%contribution would be a natural consequence of the light sigma effect.  It
%is therefore tempting to conjecture that the 
%proximity of the critical point is felt at RHIC. This conjecture can
%be tested by a direct measurement of the proton fluctuation \eq{f}. A
%serious claim, however, can be made only after a measurement of the $\sqrt s$
%dependence of the effect is performed and a correlation with other
%signatures of the critical point is observed.

% Note that if
% the end-point exists at such a low value of $\mu_B$ as considered in
% this scenario, we expect the first order phase transition at SPS
% energy. In this case the enhancement of the baryon number fluctuation
% may occur due to a quite different mechanism \cite{gavin}.


\begin{thebibliography}{99}




\bibitem{ga}
M.~Gazdzicki and S.~Mrowczynski,
%``A Method to study 'equilibration' in nucleus-nucleus collisions,''
Z.\ Phys.\ C {\bf 54}, 127 (1992);
%%CITATION = ZEPYA,C54,127;%%
M.~Gazdzicki, A.~Leonidov and G.~Roland,
%``On event-by-event fluctuations in nuclear collisions,''
Eur.\ Phys.\ J.\ C {\bf 6}, 365 (1999).
%[arXiv:hep-ph/9711422].
%%CITATION = HEP-PH 9711422;%%

\bibitem{C_V}
L.~Stodolsky,
%``Temperature fluctuations in multiparticle production,''
Phys.\ Rev.\ Lett.\  {\bf 75}, 1044 (1995);
%%CITATION = PRLTA,75,1044;%%
%\bibitem{Shuryak:1998yj}
E.~V.~Shuryak,
%``Event-per-event analysis of heavy ion collisions and thermodynamical  fluctuations,''
Phys.\ Lett.\ B {\bf 423}, 9 (1998).
%[arXiv:hep-ph/9704456].
%%CITATION = HEP-PH 9704456;%%

\bibitem{srs} M.~A.~Stephanov, K.~Rajagopal and E.~V.~Shuryak,
%``Signatures of the tricritical point in {QCD},''
Phys.\ Rev.\ Lett.\  {\bf 81}, 4816 (1998).
%[arXiv:hep-ph/9806219].
%%CITATION = HEP-PH 9806219;%%

\bibitem{srslong}
M.~A.~Stephanov, K.~Rajagopal and E.~V.~Shuryak,
%``Event-by-event fluctuations in heavy ion collisions and the {QCD}  critical point,''
Phys.\ Rev.\ D {\bf 60}, 114028 (1999).
%[arXiv:hep-ph/9903292].
%%CITATION = HEP-PH 9903292;%%

\bibitem{jk} S. Jeon and V. Koch, Phys. Rev. Lett. {\bf 83} 5435 (1999).
\bibitem{asakawa} M. Asakawa, U. Heinz and B. M\"uller, Phys. Rev. Lett. {\bf 85} 2072 (2000).

\bibitem{jeon} S. Jeon and V. Koch, Phys. Rev. Lett. {\bf 85} 2076 (2000).

\bibitem{barducci}
M.~Asakawa and K.~Yazaki,
%``Chiral Restoration At Finite Density And Temperature,''
Nucl.\ Phys.\ A {\bf 504}, 668 (1989);
%%CITATION = NUPHA,A504,668;%%
A.~Barducci, R.~Casalbuoni, S.~De Curtis, R.~Gatto and G.~Pettini,
%``Chiral Symmetry Breaking In QCD At Finite Temperature And Density,''
Phys.\ Lett.\ B {\bf 231}, 463 (1989).
%%CITATION = PHLTA,B231,463;%%


\bibitem{berges}
J.~Berges and K.~Rajagopal,
%``Color superconductivity and chiral symmetry restoration at nonzero  baryon density and temperature,''
Nucl.\ Phys.\ B {\bf 538}, 215 (1999);
%[arXiv:hep-ph/9804233];
%%CITATION = HEP-PH 9804233;%%
M.~A.~Halasz, A.~D.~Jackson, R.~E.~Shrock, M.~A.~Stephanov and J.~J.~Verbaarschot,
%``On the phase diagram of {QCD},''
Phys.\ Rev.\ D {\bf 58}, 096007 (1998).
%[arXiv:hep-ph/9804290].
%%CITATION = HEP-PH 9804290;%%

\bibitem{berdnikov} B. Berdnikov and K. Rajagopal, Phys. Rev. D {\bf 61}, 105017 (2000).

\bibitem{fodor} Z. Fodor and S. D. Katz, JHEP {\bf 0203} 014 (2002).

\bibitem{hatta} Y. Hatta and T. Ikeda, Phys. Rev. D {\bf 67}, 014028 (2003).

\bibitem{misha} 
M.~A.~Stephanov,
%``Thermal fluctuations in the interacting pion gas,''
Phys.\ Rev.\ D {\bf 65}, 096008 (2002).
%[arXiv:hep-ph/0110077].
%%CITATION = HEP-PH 0110077;%%


\bibitem{larry}
L.~D.~McLerran,
%``A Chiral Symmetry Order Parameter, The Lattice And Nucleosynthesis,''
Phys.\ Rev.\ D {\bf 36}, 3291 (1987).
%%CITATION = PHRVA,D36,3291;%%
\bibitem{kunihiro} T. Kunihiro, Phys. Lett. B {\bf 271} 395 (1991).
\bibitem{go} A. Gocksch, Phys. Rev. Lett. {\bf 67} 1701 (1991).
\bibitem{gottlieb}
S.~Gottlieb, W.~Liu, D.~Toussaint, R.~L.~Renken and R.~L.~Sugar,
%``The Quark Number Susceptibility Of High Temperature QCD,''
Phys.\ Rev.\ Lett.\  {\bf 59}, 2247 (1987).
%%CITATION = PRLTA,59,2247;%%
\bibitem{bernard}
C.~Bernard {\it et al.}  [MILC Collaboration],
%``High temperature QCD with three flavors of improved staggered quarks,''
hep-lat/0209079.
%%CITATION = HEP-LAT 0209079;%%

\bibitem{gavai} R. V. Gavai, J. Potvin and S. Sanielevici, Phys. Rev. D {\bf 40} (1989) 2743.

\bibitem{gavin} S. Gavin, nucl-th/9908070;
D.~Bower and S.~Gavin,
%``Baryon fluctuations and the QCD phase transition,''
Phys.\ Rev.\ C {\bf 64}, 051902 (2001).
%[arXiv:nucl-th/0106010].
%%CITATION = NUCL-TH 0106010;%%


\bibitem{shuryak} E. V. Shuryak and M. A. Stephanov, 
Phys. Rev. C {\bf 63} 064903 (2001).


\bibitem{reid} J.~G.~Reid  [STAR Collaboration],
%``Star Event-By-Event Fluctuations,''
Nucl.\ Phys.\ A {\bf 698}, 611 (2002).
%%CITATION = NUPHA,A698,611;%%
S. A. Voloshin [STAR Collaboration], nucl-ex/0109006; 
R. L. Ray [STAR Collaboration], Talk given at QM2002, Nantes, nucl-ex/0211030.

\bibitem{spsdata} 
S.~V.~Afanasev {\it et al.}  [NA49 Collaboration],
%``New Results From Na49,''
Nucl.\ Phys.\ A {\bf 698}, 104 (2002).
%%CITATION = NUPHA,A698,104;%%


%\bibitem{kno} Z. Koba, H. B. Nielsen, and P. Olsen, Nucl. Phys. B {\bf 40},
%317 (1972).

%\bibitem{allton} Allton et. al., Phys. Rev. D {\bf 66} (2002) 014506.


\bibitem{hatsuda} T. Hatsuda and T. Kunihiro, Phys. Lett. B {\bf 185} (1987) 304.

\bibitem{braun} P. Braun-Munzinger, K. Magestro, K. Redlich and
J. Stachel, Phys. Lett. B {\bf 518} 41 (2001).



\bibitem{sps} P. Braun-Munzinger, J. Stachel, J. P. Wessels and N. Xu,
Phys. Lett. B {\bf 365} 1 (1996).

%\bibitem{ags} P. Braun-Munzinger, J. Stachel, J. P. Wessels and N. Xu, Phys. Lett. B {\bf 344} (1995) 43.

\bibitem{zaranek} J. Zaranek, Phys. Rev. C {\bf 66}, 024905 (2002).


%\bibitem{blaizot} J. P. Blaizot, E. Iancu and A. Rebhan, Phys. Lett. B {\bf 523} (2001) 143.


\bibitem{muia}
D.~T.~Son and M.~A.~Stephanov,
%``Pion propagation near the QCD chiral phase transition,''
Phys.\ Rev.\ Lett.\  {\bf 88}, 202302 (2002).
%[arXiv:hep-ph/0111100].
%%CITATION = HEP-PH 0111100;%%

\bibitem{kogut} J.~B.~Kogut, M.~A.~Stephanov and D.~Toublan,
%``On two-color QCD with baryon chemical potential,''
Phys.\ Lett.\ B {\bf 464}, 183 (1999).
%[arXiv:hep-ph/9906346].
%%CITATION = HEP-PH 9906346;%%


%\bibitem{muI}
%B.~Klein, D.~Toublan and J.~J.~Verbaarschot,
%``The QCD phase diagram at nonzero temperature, baryon and isospin  chemical po%tentials in random matrix theory,''
%hep-ph/0301143;
%%CITATION = HEP-PH 0301143;%%
%D.~Toublan and J.~B.~Kogut,
%``Isospin chemical potential and the QCD phase diagram at nonzero  temperature %and baryon chemical potential,''
%hep-ph/0301183.
%%CITATION = HEP-PH 0301183;%%

\end{thebibliography}
\end{document}